\documentclass[12pt, a4paper]{article}
\pdfoutput=1
\usepackage{graphicx}

\usepackage{color}
\usepackage{amsmath,amssymb}

\ifx\pdfoutput\undefined
\usepackage[dvips,bookmarks=false]{hyperref}	
\else
\usepackage{hyperref}	
\fi

\hypersetup{colorlinks, citecolor=bluscuro, linkcolor=black, urlcolor=bluscuro}
\definecolor{rossos}{cmyk}{0,1,1,0.55}
\definecolor{bluscuro}{rgb}{0.15, 0.2, .85}
\definecolor{bluchiaro}{cmyk}{1,.3,0.,0.1}

\newcommand{\fref}[1]{Fig.~\ref{fig:#1}} 
\newcommand{\eref}[1]{Eq.~\eqref{eq:#1}}

\newcommand{\sref}[1]{Section~\ref{sec:#1}}
\newcommand{\cref}[1]{Chapter~\ref{ch:.#1}}

\newcommand{\nnl}{\nonumber \\}

\newcommand{\beq}{\begin{equation}} 
\newcommand{\eeq}{\end{equation}} 
\newcommand{\ba}{\begin{array}}  
\newcommand{\ea}{\end{array}} 
\newcommand{\bea}{\begin{eqnarray}}  
\newcommand{\eea}{\end{eqnarray} }  
\newcommand{\be}{\begin{eqnarray}}  
\newcommand{\ee}{\end{eqnarray} }  
\newcommand{\bal}{\begin{align}}
\newcommand{\eal}{\end{align}}   
\newcommand{\bi}{\begin{itemize}}  
\newcommand{\ei}{\end{itemize}}  
\newcommand{\ben}{\begin{enumerate}}  
\newcommand{\een}{\end{enumerate}}  
\newcommand{\bc}{\begin{center}}
\newcommand{\ec}{\end{center}} 
\newcommand{\bt}{\begin{table}}
\newcommand{\et}{\end{table}}  
\newcommand{\btb}{\begin{tabular}}
\newcommand{\etb}{\end{tabular}}  
\newcommand{\bvec}{\left ( \ba{c}}
\newcommand{\evec}{\ea \right )}

\newcommand{\cO}{{\mathcal O}} 
 
\newcommand{\cL}{{\mathcal L}} 
 
\newcommand{\cM}{{\mathcal M}}

\def\hc{{\rm h.c.}} 

\def\lsim{\mathrel{\rlap{\lower3pt\hbox{\hskip0pt$\sim$}}
   \raise1pt\hbox{$<$}}}         
\def\gsim{\mathrel{\rlap{\lower4pt\hbox{\hskip1pt$\sim$}}
   \raise1pt\hbox{$>$}}}         

\begin{document}

\begin{titlepage}

\vspace*{-2cm}
\begin{flushright}
LPT Orsay 19-05 \\
\vspace*{2mm}
\end{flushright}

\begin{center}

\vspace{1cm}
{\LARGE Which EFT} 
\vspace{1cm}

\renewcommand{\thefootnote}{\fnsymbol{footnote}}
{\bf  Adam Falkowski$^a$ and Riccardo Rattazzi$^b$}
\renewcommand{\thefootnote}{\arabic{footnote}}
\setcounter{footnote}{0}

\vspace*{.5cm}
\centerline{${}^a$\it  Laboratoire de Physique Th\'{e}orique (UMR8627), CNRS, Univ. Paris-Sud,}
\centerline{\it Universit\'{e} Paris-Saclay, 91405 Orsay, France}\vspace{1.3mm}
\centerline{$^b$ \it Theoretical Particle Physics Laboratory (LPTP), Institute of Physics,}
\centerline{EPFL, Lausanne, Switzerland}
\vspace*{.2cm}

\end{center}

\vspace*{5mm}
\begin{abstract}\noindent\normalsize

We classify  effective field theory (EFT) deformations of the Standard Model (SM) according to the analyticity property of the Lagrangian as a function of the Higgs doublet $H$. Our distinction in  {\em analytic} and {\em non-analytic} 
corresponds to the more familiar one  between linearly and non-linearly realized electroweak symmetry, but offers  deeper physical insight. 
From the UV perspective, non-analyticity occurs when the new states acquire mass from electroweak symmetry breaking, and thus cannot be decoupled to arbitrarily high scales.  
This is reflected in the IR by the anomalous growth of the interaction strength for processes involving many Higgs bosons and  longitudinally polarized massive vectors, with  a  breakdown of the EFT description below a scale $\cO(4 \pi v)$.
Conversely,  analyticity occurs when new physics can be pushed 
 parametrically above  the electroweak scale.   

We illustrate the physical distinction between these two EFT families by discussing  Higgs boson self-interactions. 
In the analytic case, at the price of some un-naturalness in the Higgs potential,  there exists space for $\cO(1)$ deviations of the cubic coupling,  compatible with  single Higgs and electroweak  precision measurements,  and with new particles out of the direct LHC reach.   Larger deviations are possible, but  subject to less robust assumptions about higher-dimensional operators in the Higgs potential. 
On the other hand, when the cubic coupling is produced by a non-analytic deformation of the SM, 
 we show by an explicit calculation that the theory reaches strong coupling at  $\cO(4 \pi v)$, quite independently of the magnitude of the cubic enhancement.  

\end{abstract}

\end{titlepage}
\newpage


\section{Introduction}  

According to the modern Wilsonian viewpoint, any Quantum Field Theory (QFT) should  be viewed as an effective description  valid below some physical energy cut-off scale. 
From this perspective, renormalizable QFT is but a useful idealization where the UV cut-off scale is either exponentially large, at least at weak coupling, or even infinite, in the case of asymptotically free theories. 
The Standard Model (SM), when limited to its renormalizable interactions, can indeed be extrapolated to energy scales of the order of the Planck scale, raising the conceptual possibility that the next  layer in particle physics be at such ultrashort distances. 
Whether that is the case or not,  it is quite certain that the effective description of physics at lower energies will not be limited to the few renormalizable couplings of the SM. 
We expect a much richer structure  deforming the leading renormalizable SM through an infinite set of non-renormalizable interactions.
The lack of direct evidence of new physics at the LHC has indeed boosted the relevance of indirect searches for such deformations.
Along these lines, many authors have pursued a variety of effective field theory (EFT) extensions of the SM.  
 Those relevant for the Higgs sector are particularly motivated in view of the well known conceptual problems associated with the existence of an elementary scalar particle.  
This paper makes a simple observation, which  provides a sharp structural classification of these EFTs.

In the construction of effective theories, symmetries play a central role. 
For instance, in the very case of EFTs for the Higgs sector, flavor symmetries are obviously crucial to tame flavor changing neutral currents. 
The role of gauge symmetries is perhaps more subtle, as they mostly  control the strength of the interaction and  the range of validity of the EFT. 
Our main point, which  concerns precisely these aspects,  can be summarized as follows. 
The most general EFT deformation of the SM Higgs sector is given by a general lagrangian invariant under the color and electromagnetic  $SU(3)_C \times U(1)_Q$ symmetry that couples the Higgs boson $h$ to other SM fields. 
To carry out this construction there  is no need whatsoever for manifest $SU(2)_W \times U(1)_Y$ electroweak (EW)  gauge invariance,
as in the broken theory one can always pick the unitary gauge. 
But unitary gauge, while making the particle content explicit, makes the structure of interactions less transparent. 
Indeed our sharp structural classification of EFTs 
is most succinctly formulated when the triplet of  Goldstone bosons $\pi_i$ eaten by $W$ and $Z$ is kept manifest so as 
to  form, together with $h$, a doublet $H$ transforming linearly under  $SU(2)_W \times U(1)_Y$:
\beq
\label{eq:H}
H\equiv  { 1 \over \sqrt 2}  e^{i\pi_i \sigma^i}\bvec 0 \\  v + h \evec.  
\eeq
We stress that, whatever the origin of $h$, we can always form such a linear multiplet. 
Two possibilities are then given for the lagrangian as a function of $H$: {\em it is either analytic or non-analytic at $H=0$}. 
More precisely: either the lagrangian is analytic, possibly after a field redefinition, or there is no field redefinition that renders it analytic.
The distinction between these two possibilities is not  aesthetic but purely dynamical. 
In the analytic case the lagrangian is polynomial in all fields, $H$ included. 
This is the more familiar case, where small deviations from the renormalizable SM are compatible with a large cut-off scale.
More technically, the ultimate cut-off grows like an inverse power of the size of the deviation. 
The situation is sharply different in the non-analytic case. 
There, as we shall illustrate in detail, the cut-off basically reduces to $\cO(4 \pi v) \sim 3$ TeV, 
where $v \approx 246$~GeV denotes the vacuum expectation value (VEV) of $H$ in \eref{H}. 
Even when deviations in the single or double Higgs production happen to be small, the low cut-off  will become manifest in processes involving many Higgs bosons and longitudinally polarized massive vector bosons.
In hindsight this result has a simple interpretation from a top-down perspective. 
The singularity at $H=0$, signaling the breakdown of the EFT,  must be associated to some heavy degree of freedom becoming massless at $H=0$. 
In other words, non-analytic EFTs simply correspond to the presence of new massive states whose mass is fully controlled by the Higgs VEV.  
The familiar relation between coupling and mass $m_*\sim g_* v$, together with the naive dimensional analysis (NDA) expectation $g_*\lsim4\pi$ immediately imply the upper bound $4\pi v$ for the mass defining the UV cut-off. 
Our distinction between analytic and non-analytic lagrangians coincides with the distinction, in use in the  Higgs EFT community,  between linear (so-called SMEFT)  and non-linear (so-called HEFT)  effective theory, or equivalently between $h$ being or not being part of a $SU(2)_W$ doublet. 
We however believe our classification is more adequate and enlightening from a physical point of view.

The classification we advertise is generally applicable to EFT extensions of the SM.
In this paper we shall illustrate it in the specific case of the Higgs potential.  
That will allow us to make the discussion very concrete and focused.
As a bonus, we will derive useful results relevant for the ongoing explorations of the cubic Higgs self-coupling.

\vskip 1.0truecm

The interest in measuring Higgs self-interactions is fueled by the hope that it may contain a clue about the more fundamental theory underlying the SM.  
Indeed, the Higgs potential is arguably the most ad-hoc element of the SM, 
and it is reasonable to suspect that the true dynamics driving  the Higgs field to acquire a  VEV is described by a more sophisticated scenario.
The current efforts are mostly focused on the cubic self-coupling. 
The coefficient $\lambda_3$ of the $h^3$ term in the SM lagrangian is completely determined by two precisely measured observables: the Higgs boson mass and the Fermi constant. 
While many other SM predictions in the Higgs sector have been successfully tested with $\cO(10 \%)$ accuracy~\cite{Khachatryan:2016vau}, probing the Higgs self-interactions remains challenging. 
The ongoing experimental effort in this direction consist in measuring the Higgs boson pair production rate~\cite{Aad:2019uzh,Sirunyan:2018two},  which is sensitive at tree level to $\lambda_3$.
In parallel, the cubic can be constrained through its one-loop effects~\cite{McCullough:2013rea} in single Higgs production at the LHC~\cite{Gorbahn:2016uoy,Degrassi:2016wml,DiVita:2017eyz,Maltoni:2017ims} and in  EW precision measurements \cite{Degrassi:2017ucl,Kribs:2017znd},  or through tree-level effects in single Higgs production in association with two W/Z bosons~\cite{Henning:2018kys}.
However, all of these methods currently leave room for a large $\cO(10)$ deviation of $\lambda_3$ relative to the SM prediction. 

This paper discusses the range of the Higgs cubic coupling  that can be generated by a dynamics beyond the SM (BSM).
The analysis depends on whether the Higgs potential at energy scales below $m_* \gg m_h$ is an analytic or non-analytic function of $H^\dagger H$. 
We start with the former case in \sref{SMEFT}.  
This case is equivalent to the so-called SMEFT~\cite{Buchmuller:1985jz,Leung:1984ni}, where various terms in the  lagrangian are organized according to their canonical dimensions, with dimension $D$ terms suppressed by $m_*^{D-4}$ powers of the BSM scale.  
 We review the power counting that controls the coefficients of various terms in the potential, stability conditions, 
  and phenomenological constraints on these coefficients from the LHC measurements of the Higgs mass and couplings. 
We are interested in a phenomenologically viable scenario where   1) $m_*$ is much bigger than $m_h$ and outside the LHC reach, and 2)  the magnitude $\xi$ of relative BSM corrections to single Higgs couplings satisfies the LHC bounds $\xi \lesssim 0.1$.  
In this setting corrections to the Higgs cubic are generated at the level of dimension-6 operators in the Lagrangian. 
We demonstrate that the cubic enhancement $\Delta_3 \equiv  {\lambda_3 \over \lambda_{3, \rm SM} } -1$  can be as large as $\cO(1)$ when the coupling strength $g_*$ in the UV theory at $m_*$ is moderately strong. 
Remarkably, $\Delta_3$ can largely exceed the relative corrections to single Higgs couplings. 
This can be understood by noting that $\lambda_3$ is a relevant coupling that becomes strong when $m_h\to 0$, with the cubic coefficient in the potential held fixed. 
More precisely we find that the cubic enhancement in the range  
 \beq
 \label{eq:bound}
 0 \lesssim \Delta_3  \lesssim  2
 \eeq 
is possible for $g_*$ moderately strong and generic coefficients of higher-dimensional operators in the Higgs potential. 
Larger or negative corrections are possible, but are subject to more stringent assumptions in order to ensure vacuum stability.     
Overall, we find  $|\Delta_3| \lesssim 4$ can be obtained for a reasonable hypothesis  about dimension-8 operators in the Higgs potential.

In \sref{HEFT} we relax the assumption that the scalar potential is a polynomial or analytic function of  $H^\dagger H$.
It is possible to add to the SM lagrangian terms of the form $\left (H^\dagger H \right )^{n/2}$ with integer $n$, 
which in the unitary gauge yield Higgs boson self-interactions $h^k$ with $k \leq n$.  
In particular, we can arrange such non-analytic terms  to contribute to  $\Delta_3$, with or without  affecting other Higgs (self-)interaction terms.   
An EFT lagrangian that has the SM local symmetry and degrees of freedom but is non-analytic in  $H^\dagger H$ is equivalent to the so-called HEFT framework (which is usually formulated without introducing the Higgs doublet field $H$,  using the language of a non-linearly realized EW symmetry, see e.g. Sec.~II.2.4 of~\cite{deFlorian:2016spz} for a review).
This framework naively offers more freedom to arrange for a large cubic Higgs coupling without violating theoretical and phenomenological bounds.  
We will argue however that in the presence of the non-analytic terms  it is impossible to parametrically separate $m_*$ and $v$, 
and instead new degrees of freedom must appear at $m_* \lesssim 4 \pi v$. 
Technically, this happens due to the wrong  (inconsistent with perturbative unitarity) behavior of the tree-level amplitudes of the form 
\beq
\label{eq:nhiggsamp}
\cM(\underbrace{V_L \dots V_L}_m \underbrace{h \dots  h}_n ),  
\eeq 
where $m \geq 2$, and  $n \geq 3$, and $V_L$ stands for longitudinally polarized $W$ or $Z$ bosons.   
That conclusion depends very weakly (logarithmically) on the magnitude of the non-analytic deformation; 
in fact, the amplitudes in \eref{nhiggsamp} hit strong coupling at $\cO(4\pi v)$ even when non-analytic terms generate a relatively small correction to the cubic term,  $|\Delta_3| \ll 1$.   
We conclude that the presence of non-analytic terms in the Higgs potential leads to  $m_* \lesssim 4 \pi v$ and typically $\xi \sim 1$, contrary to the assumptions of this analysis. 
 The wrong behavior of the amplitudes in \eref{nhiggsamp} can be controlled only when the Higgs potential is well-approximated by a polynomial in $H^\dagger H$.  
This however brings us back to the  SMEFT case and to the bound in \eref{bound}.

\section{Analytic Higgs potential (SMEFT)}
\label{sec:SMEFT}

Let us first define our notation and introduce the relevant physical quantities. 
In complete generality, the potential for the Higgs boson field $h$ takes the form
\beq
V(h) = {m_h^2 \over 2} h^2  + {m_h \over 3!} \lambda_3 h^3 +  {1 \over 4! } \lambda_4 h^4 + \sum_{n=5}^\infty {\lambda_n \over n! m_h^{n-4}} h^n  . 
\eeq 
In the SM this arises by expanding around the vacuum the potential
\beq 
V_R(H^\dagger H)=- \frac {m_h^2}{2} H^\dagger H +\frac{\lambda_h}{4} (H^\dagger H)^2\qquad \qquad \lambda_h\equiv 2m_h^2/v^2\, .
\label{VSM}
\eeq
The SM cubic and quartic couplings take values  
$\lambda_3 = \lambda_{3, \rm SM} \equiv {3 m_h \over  v}=3\sqrt {\lambda_h\over 2}$,  
$\lambda_4 = \lambda_{4, \rm SM} \equiv {3 m_h^2 \over  v^2}={3\lambda_h\over 2}$, while $\lambda_{n>4}$ vanish.
Our goal is to set a theoretical bound on the relative deformation $\Delta_3 \equiv 
\lambda_3/\lambda_{3, \rm SM} - 1$ of the cubic coupling.  In the SM the observed values of $m_h$ and $v$ imply $\lambda_h\simeq 1/2$, which is well within the perturbative regime. Indeed standard estimates of the perturbative upper bound of $\lambda_h$  range roughly between $3\pi^2$ and  $10\pi^2$ in accordance with NDA \footnote{%
More precisely the RG evolution estimate used in \cite{Contino:2017moj} suggests the lower of the values, while the scattering phase method of \cite{Glioti:2018roy} yields the upper value.}. 
Choosing for definiteness a reference strong coupling value $\bar \lambda_h \equiv 4\pi^2$ we have $\lambda_h/\bar \lambda_h\sim 0.01$. The SM quartic 
is thus about two orders of magnitude below its perturbative upper bound, while the cubic 
is accordingly about  one order of magnitude below its perturbative upper bound.
A fair question is what portion of this range can be covered by plausible extensions of the SM.

In this section we tackle this question in the framework of the SMEFT with higher-dimensional operators.\footnote{%
See also Ref.~\cite{DiLuzio:2017tfn}. 
Our analysis offers a different perspective, emphasizing the dependence on the microscopic properties of the UV theory and fine-tunings required by phenomenology.  
Moreover we include in our discussion the impact  of $D\leq 8$ operators on the stability of the Higgs potential. }
Consider the SMEFT arising as a low-energy approximation of a microscopic theory with  fundamental scale $m_*$ and maximal coupling size $g_*$, focussing in particular on the Higgs potential. 
It is also convenient to define $f\equiv {\sqrt 2}m_*/g_*$.
We will assume $m_* \gg m_h$, in which case one can organize the SMEFT operators in a meaningful expansion in $1/m_*$, 
and estimate the size of various Wilson coefficients using the usual power counting rules~\cite{Giudice:2007fh,Liu:2016idz,Contino:2016jqw}.
Assuming the existence of a minimum at $\langle H^\dagger H\rangle \equiv v^2/2$,  the potential has the general form
\beq
\label{eq:VHsmeft}
V(H)=\frac{m_*^4}{4 g_*^2} \sum_{n\geq 2}^{\infty}a_n X^n\equiv \frac{m_*^4}{4 g_*^2} X^2 P(X), 
\qquad\qquad X\equiv {2H^\dagger H-v^2 \over f^2} , 
\eeq
with $a_n\leq \cO(1)$.
The upper bound on the $a_n$ coefficients corresponds to the absence of couplings stronger than $g_*$ at the scale $m_*$.
Some couplings could consistently, naturally or unnaturally, be tuned to be small. For instance in the simplest instance of composite pseudo-Nambu-Goldstone Higgs we have $a_n\sim y_t^2/16\pi^2$ for any $n$, where $y_t$ is the top Yukawa coupling.
On the other hand for an ordinary  scalar  in a generic theory characterized by $g_*$ and  $m_*$ we would  expect $a_n=\cO(1)$ and $v \sim f$. But such a generic theory is at odds with phenomenology and some tuning is always necessary. Consider first the relation $v\sim f$.  Indeed,
defining $\xi \equiv v^2/f^2$, several independent dimension-6 SMEFT operators, 
such as e.g. $({\partial_\mu} |H|^2)^2$ or $|H|^2 \bar t_R H Q_3$,   would produce deformations of single Higgs couplings of relative size  $\cO(\xi)$. 
In view of the agreement of the LHC Higgs data with the SM predictions  we will thus assume  $\xi \lesssim 0.1$ in the discussion below.
In concrete models the relation $\xi \ll 1$ is typically achieved by fine tuning. This single tuning of $\xi$ appears more plausible than the tuning of multiple coefficients required to match Higgs data in a theory with $\xi =\cO(1)$.
  
Another independent tuning may be needed to ensure that the Higgs boson mass $m_h$ matches the observed value.  
Eq.~(\ref{eq:VHsmeft}) implies $m_h^2 = a_2 m_*^2 \xi^2  =   {a_2 \over 2} g_*^2 v^2 $, so that according to the definition of $\lambda_h$ in eq.~(\ref{VSM}) we can write 
\beq 
\lambda_h^2 =a_2 g_*^2 \, .
\eeq  
This shows that,  when the UV coupling $g_*$ is strong, a tuning of order $\lambda_h/g_*^2$ for $a_2$ is needed.
The strongest tuning,  $a_2 \sim  0.01$, corresponds to the case $g_*\sim \sqrt {\bar  \lambda_h}=2\pi$ in which a generic $a_2=1$ would produce a maximally strong $\lambda_h$. In view of these properties this scenario was referred to as an {\it accidentally light Higgs} in Ref.~\cite{Liu:2016idz}.

Before proceeding we would like to make a little digression concerning the naive expectation $a_n\sim \cO(1)$ in a generic theory.
Indeed one should be more careful especially when considering $n\gg 1$, corresponding to operators with many legs. 
It would be nice to have the analogue of NDA including an estimate for the scaling with $n$. 
We cannot offer a general self-consistent analysis along these lines, but we can discuss a few simple models, where \eref{VHsmeft} is generated by either tree or one-loop graphs. 
One finds the rough scaling $a_n\sim k^n n^{-\alpha}$, with $k$ and $\alpha$ depending on the model. 
Now, the factor $k^n$ simply corresponds to the ambiguity in the definition of $g_*$. Indeed a redefinition $g_*\to g_* k$ implies precisely the redefinition $a_n\to a_n k^{n-2}$, which up to a constant coefficient produces the same scaling. 
The power-law dependence on $n$ is  more structural. 
In our experience $\alpha$ can range from $0$ to $5/2$. In particular, a simple UV model with potential $V=m_*^3/g_* \phi +(m_*^2+g_*^2 H^\dagger H) \phi^2$ produces, upon integrating out $\phi$, a series with $\alpha =0$. 
This result simply follows from the geometrical series generated by the $\phi$ propagator. On the other hand, other UV variants like 
$V=g_*m_*H^\dagger H \phi +m_*^2 \phi^2 +g_*m_* \phi^3 +g_*^2 \phi^4$, produce a series with $\alpha=5/2$. 
In both cases the scaling of $a_n$ is consistent with the breakdown of the low-energy expansion for $g_*^2 |H|^2 \sim m_*^2$, which is physically expected.

Expanding $V(H)$ around its minimum at $\langle H^\dagger H\rangle \equiv v^2/2$ (i.e. $X=0$), we readily obtain the low-energy 
Higgs self-couplings. In particular for the cubic and quartic we find
\begin{eqnarray}
\label{cubic}
\lambda_3& =&
\frac{3}{\sqrt 2} g_*\left ( \sqrt a_2+\frac{2 a_3}{\sqrt a_2} \xi\right ) 
=\lambda_{3, \rm SM}  \left (1+2 \frac{a_3}{a_2} \xi\right ), \qquad   
 \lambda_{3, \rm SM}  = \frac{3 \sqrt a_2}{\sqrt 2}   g_* ,
 \nnl 
\lambda_4& = &
 \frac{3}{2} g_*^2\left ( a_2+12 a_3\xi+16 a_4\xi^2\right ) 
= \lambda_{4, \rm SM} \left(
1+12 \frac{a_3}{a_2}\xi+16 \frac{a_4}{a_2}\xi^2\right ), \qquad    \lambda_{4, \rm SM}  = \frac{3 a_2}{2}  g_*^2 , 
\nnl 
\end{eqnarray}
where we factored out the SM result obtained in the limit $a_{n>2}=0$. 
These expressions show that for $a_2\ll 1$ (thus for $g_*$ moderately strong) one can obtain sizable deviations from the SM even for relatively small $\xi$. More specifically, by considering the couplings written in terms of $g_*$, one sees that, within the range $a_2,\xi\ll 1$,  $a_3, a_4 \lsim \cO(1)$, one can choose $a_3\xi/\sqrt a_2=\cO(1)$ so as to enhance $\lambda_3$ up to $\cO(g_*)$. 
This happens because the cubic coupling is {\it relevant} and becomes strong when $m_h\to 0$, with the cubic coefficient in the potential  held fixed. 
 Numerically, the correction to the cubic Higgs coupling relative to the SM one is given by 
\beq
 \Delta_3 \approx  20 a_3 \left ( {0.01 \over a_2} \right ) \left ( \xi \over 0.1 \right ) ,
\eeq  
and naively it can be larger than $\cO(10)$ for a sufficiently strong coupling in the UV theory.\footnote{%
Note that for $\xi \ll 1$ the relative corrections to $\lambda_3$ and to $\lambda_4$ are both of order $a_3\xi/a_2$,
which implies that in principle the two approach the strong coupling differently. 
However, phenomenological constraints and numerical factors disturb this NDA, and as a result the respective strongly coupled values,  $\lambda_3 \approx \sqrt{8 \pi^2}$ and $\lambda_4 \approx 8 \pi^2$,  are reached  more or less simultaneously as $a_3\xi/a_2$ is increased. }  
The above conclusion, however, does not take into account the  requirement of absolute stability of the EW vacuum. 
 Indeed it is obvious\footnote{%
 Nevertheless this was overlooked in Ref.~\cite{Liu:2016idz}. } 
 that, keeping all other terms fixed, the coefficient of $h^3$ cannot be made arbitrarily large without generating a second minimum deeper than the one at $h=0$. 
In the following we quantify the stability constraints. 

Our potential has the form $V\propto X^2 P(X)$ with $P(X)= a_2+a_3 X+a_4X^2+\dots $ for $X\in [-\xi, +\infty)$, with $X=-\xi$ corresponding to the EW preserving vacuum $\langle H^\dagger H\rangle=0$. 
For $a_2>0$ we have a realistic local minimum at  $X=0$, where $V$ vanishes. 
Unless this minimum is also global, it will be destabilized by vacuum tunneling.
The condition for metastability thus basically coincides with the condition for absolute stability: $P(X)\geq 0$ for $X\in [-\xi, +\infty)$. 
In order to make the discussion more transparent it is convenient to work with the rescaled variable $\tilde X=X/\xi$, which is defined in the domain $ [-1, +\infty)$.  Writing $P(X)=a_2 \tilde P(\tilde X)$ we have 
\beq
\label{eq:tildeP}
\tilde P(\tilde X)=  1+c_3 \tilde X+c_4{\tilde X}^2+\dots  \qquad{\mathrm {with}}\quad c_n\equiv \frac{a_n}{a_2}\xi^{n-2}. 
\eeq
The coefficient $c_3$ of the linear term  is directly related to the correction to the cubic coupling in Eq.~(\ref{cubic}): 
 $\Delta_3= 2 c_3$, while  $c_{n>3}$ encode effects of dimension-8 and higher SMEFT operators in the Higgs potential. 
Now, under the assumption $a_n\lsim \cO(1)$, the experimental constraints $\xi\lsim 0.1$ and $a_2\gsim 0.01$   imply 
\bea  
\label{ccoeff}
 |c_3|  & =&    { |a_3| \over a_2 }  \xi    =  10 \left ( 0.01 \over a_2 \right ) \left ( \xi \over 0.1 \right )  \lesssim  \cO(10), 
\nnl 
 |c_4|  & = & {|a_4| \over a_2} \xi^2  =  \left ( 0.01 \over a_2 \right )  \left ( \xi \over 0.1 \right )^2  \lesssim  \cO(1), 
 \nnl 
|c_{ n>4}| & = &  {|a_n| \over a_2} \xi^{n-2} =  \left ( 0.01 \over a_2 \right )  \left ( \xi \over 0.1 \right )^2  \xi^{n-4}    \ll \cO(1) . 
\eea 
We conclude that for $\xi \ll 1$ the parameters  $|c_{n>3}|$ are suppressed with respect to $|c_3|$. 
It is now clear why,  for large $|\Delta_3|$, stability is an issue. 
The behavior of the potential at small  $\tilde X$ is dominated by the first two terms in \eref{tildeP}. 
It follows that for $|c_3| \gg 1$ the function $\tilde P$ will cross zero near the origin at $\tilde X\simeq \tilde X_c \equiv -1/c_3$, {\it i.e.} within the physical domain $ [-1, +\infty)$, leading to a deeper minimum of $V(H)$ than the one at $\langle H^\dagger H \rangle = v^2/2$.
Thus, the correction to the Higgs cubic coupling larger than $\cO(1)$ may lead to an instability.

\begin{figure}[h]
\begin{center}
\includegraphics[width=0.48\textwidth]{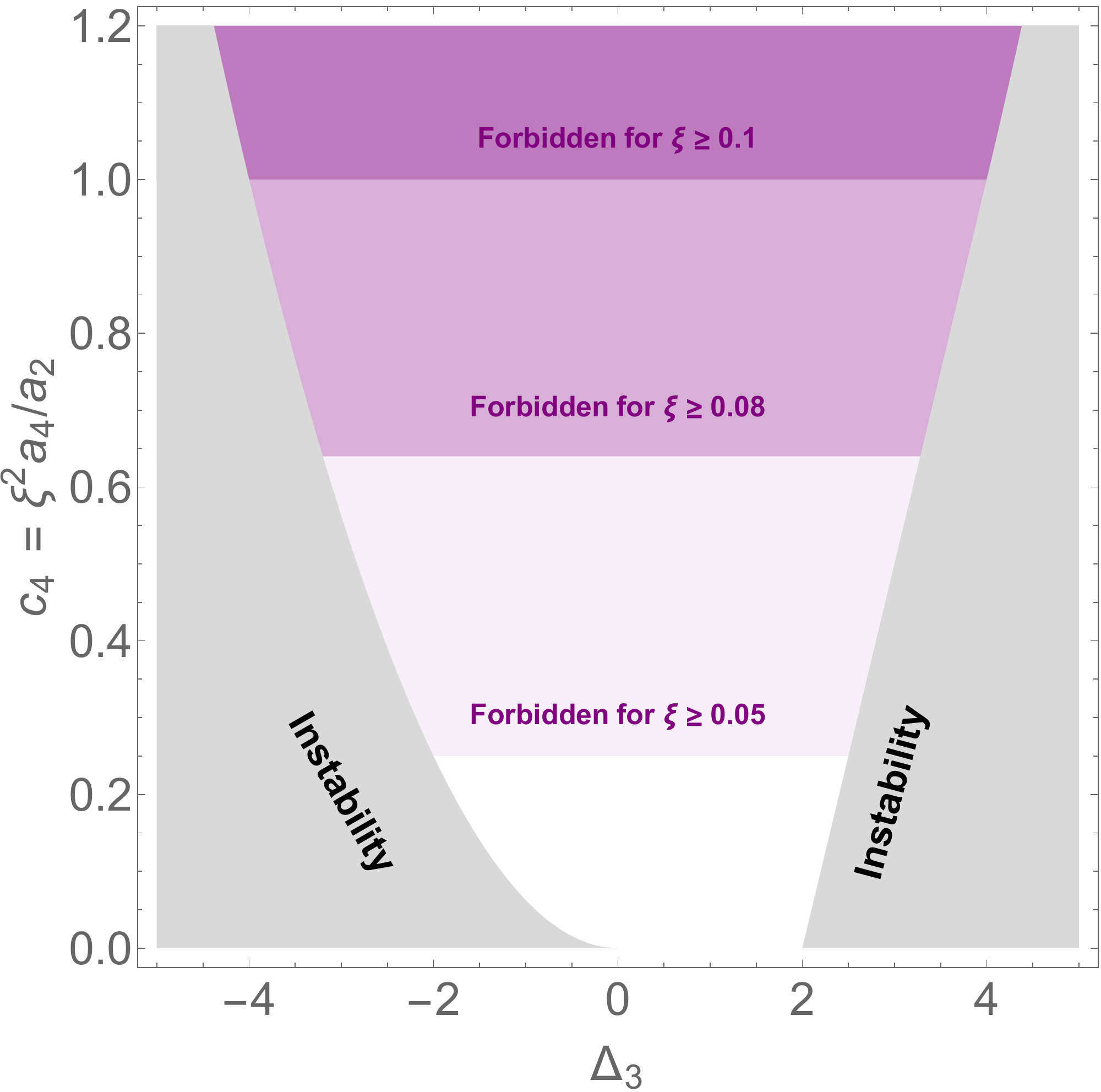}
\includegraphics[width=0.48\textwidth]{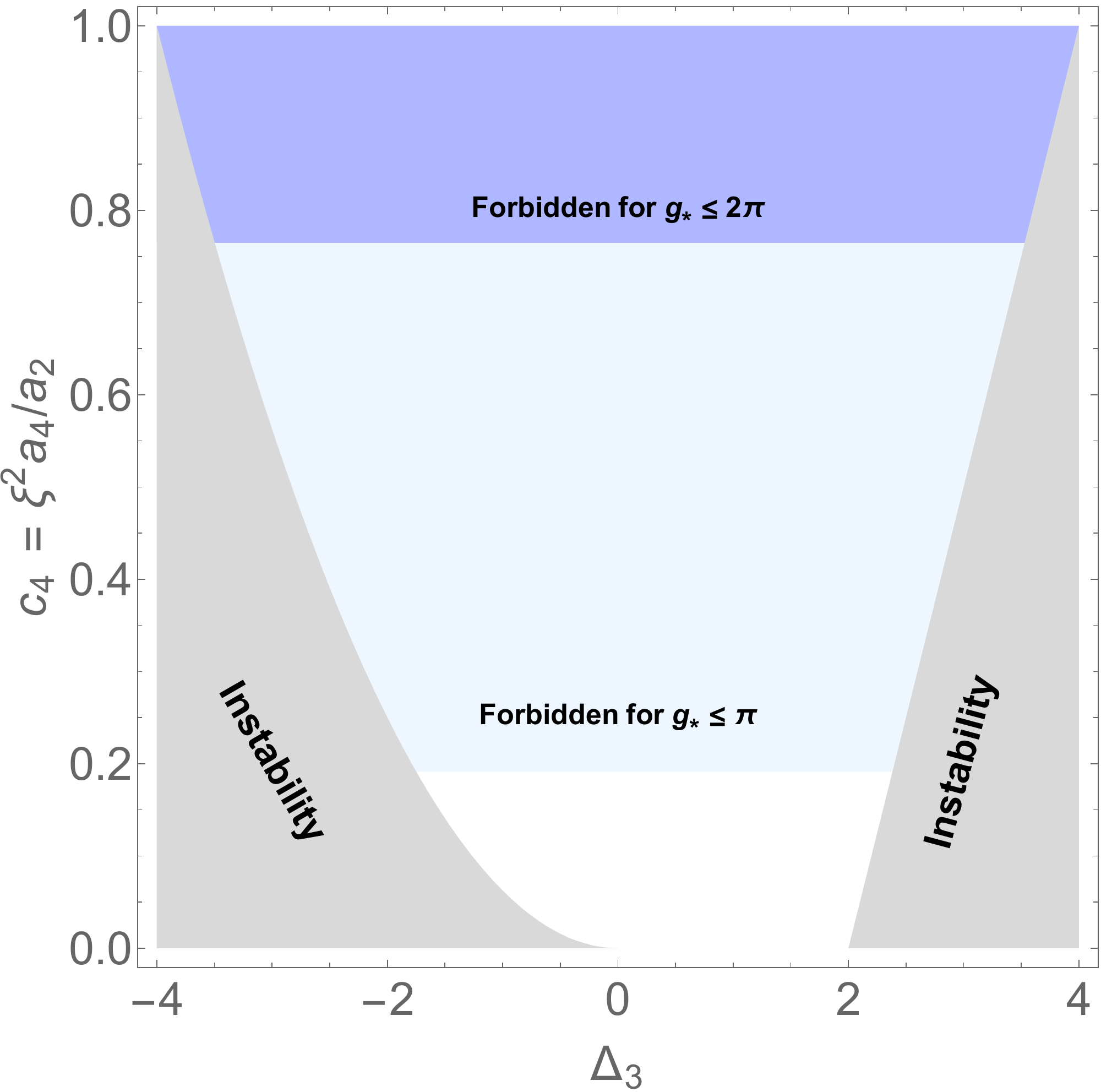}
\caption{
\label{fig:analytic}
Parameter space for the cubic Higgs self-coupling deformation $\Delta_3$ relative to the SM value. 
The allowed region depends on the value $c_4 = \xi a_4/a_2$, which encodes effects of dimension-8 SMEFT operators in the Higgs potential. 
The gray area is excluded by stability considerations, as the potential contains  a deeper minimum that the EW vacuum 
at $\langle H^\dagger  H \rangle = v^2/2$. 
Left: the purple areas are excluded for $a_4 =1$ and $a_2 = 0.01$ under different hypotheses about the parameter $\xi = v^2/f^2$, which characterizes the size of the corrections to the single Higgs boson couplings to matter.  
Right: the blue areas are excluded for $a_4=1$ and $\xi = 0.1$ under different hypotheses about the coupling strength $g_*$ of the BSM theory underlying the SM. 
}
\end{center}
\end{figure}

To make the bound more precise, it is quantitatively adequate to focus on the case 
\beq 
\tilde P =  1+{1 \over 2} \Delta_3  \tilde X+c_4 {\tilde X}^2, 
\eeq 
given that the $|c_{ n>4}|$ are anyway expected to be  suppressed. 
The resulting constraints are shown in \fref{analytic}. 
Outside the region $0<\Delta_3 < 4$ the bound coincides
with the condition for  absolute positivity of $\tilde P$: $\Delta_3^2 < 16 c_4$.
Using the  definition of $c_4$  in \eref{tildeP} we obtain 
\beq 
\label{eq:deltalambda3_inequality}
|\Delta_3| \lesssim 4 \sqrt{a_4} \sqrt{ 0.01 \over a_2 } \left ( \xi \over 0.1 \right ).  
\eeq 
On the other hand, for $0<\Delta_3 <4$, the bound is weaker corresponding to cases where $\tilde P$ becomes negative in the unphysical region $\tilde X<-1$. 
In particular, for $0\leq \Delta_3 \leq 2$, $c_4$ is even allowed to vanish. 
All in all, we conclude that a correction to the Higgs cubic coupling in the  range 
$0\leq \Delta_3 \leq 2$ can be obtained under the very conservative assumption  $a_4>0$. 
Larger or smaller values of $\Delta_3$ are possible, subject to assumptions about the coefficient $a_4$   such that \eref{deltalambda3_inequality} is satisfied. In particular  $|\Delta_3| > 4$  can only be achieved for 
 $a_4 \gsim \cO(1)$ which seems less plausible. 
The maximal value is reached for a maximally strongly coupled BSM theory completing the SMEFT at the scale $m_*$.  
For more moderate (and perhaps more realistic) couplings, the bound is correspondingly stronger; 
for example under the condition $|a_4|<1$ one has $-1.7 \lesssim \Delta_3 \lesssim 2.4$ for $g_* = \pi$. 
Furthermore, as illustrated in  \fref{analytic}, the bound can be tightened if the {\em single} Higgs couplings to matter are better constrained by experiment, leading to a stronger bound on the parameter $\xi$. 
Notice however that the region $0\leq \Delta_3 \leq 2$ can still be covered even for relatively weak couplings. For instance  for $a_3=1$, $a_2=0.1$ and $\xi = 0.1$
one can reach up to $\Delta_3=2$. 

One way to read our results is that there exists space for a strongly coupled accidentally light Higgs  with sizable $\cO(1)$
deviations in its self-couplings but compatible
with all single Higgs and  EW precision measurements ($\xi \lsim 0.1$) and with a fundamental scale $m_*\sim g_* v/\sqrt \xi\lsim 5 $ TeV
out of reach of present LHC direct searches.

\section{Non-analytic Higgs potential (HEFT)}
\label{sec:HEFT}

In the previous section we have demonstrated that, in the SMEFT framework with the parametric separation between BSM and EW scales,  theoretical arguments and experimental constraints lead to an upper bound on the magnitude of the cubic self-coupling of the Higgs boson:  $|\Delta_3| \lesssim $~few.  In particular we showed that,  at the price of a tuning of $m_h^2$ and $v^2$, an $\cO(1)$ deviation $\Delta_3$ can be obtained consistently with present data  and for a new physics scale $m_*$
 above the present LHC reach. Essential in the derivation was the analytic dependence of the lagrangian on the Higgs doublet field $H$, which follows from the assumption that the heavy states are massive regardless of EW symmetry breaking.
In this section we discuss the cubic self-coupling in a setting where the analyticity assumption is removed. 
We shall see that, going beyond the SMEFT, there is an obstruction to achieving the separation between the BSM and EW scales.
This other scenario is therefore subject to much more severe constraints coming from direct and indirect searches for new physics.
 
Consider, for concreteness, a simple scenario of an EFT where the Higgs boson self-interactions are described by the potential 
\beq
\label{eq:TRIPLE_hself}
V(h)  =   {m_h^2 \over 2 v} \left ( 1 + \Delta_3 \right ) h^3   + {m_h^2 \over 8 v^2} h^4.  
\eeq  
where the only deviation from the   SM resides in the cubic coupling. In particular all $h^n$ terms with $n \geq 5$ are absent.
Note that such a pattern  cannot be obtained from any $SU(2)_W \times U(1)_Y$  invariant  potential that is an analytic function of $H^\dagger H$.  
In particular, it cannot be obtained in the SMEFT, unless the entire infinite tower of higher-dimensional operators contributes to the potential. 
That situation however corresponds to $f\sim v$, which is phenomenologically very implausible.  
On the other hand,  \eref{TRIPLE_hself}  belongs to the parameter space of the so-called HEFT, which is an effective theory where only the $U(1)_Q$ part of the EW symmetry is linearly realized. 
In the HEFT, the Goldstone bosons eaten by $W$ and $Z$ transform non-linearly under the full EW symmetry, while the Higgs boson $h$ is a perfect singlet. 
As a consequence, the general potential $V = v^4 \sum_{n = 2}^\infty c_n (h/v)^n$ with arbitrary coefficients $c_n$ is allowed by the symmetries, and \eref{TRIPLE_hself} represents one particular direction within the HEFT parameter space. 

It is illuminating to rewrite \eref{TRIPLE_hself} in a manifestly  $SU(2)_W \times U(1)_Y$ invariant language: 
\beq
\label{eq:TRIPLE_hnonanal}
V(H)  = {m_h^2 \over 8 v^2} \left (2 H^\dagger H - v^2 \right )^2  + \Delta_3  {m_h^2 \over 2 v} \left (\sqrt{2 H^\dagger H} - v \right )^3 , 
\eeq  
where $H$ is the Higgs field in \eref{H}. 
In the unitary gauge, $\pi_i  =0$,   this potential reduces \eref{TRIPLE_hself}. 
We should mention that we are not aware of a concrete UV-complete model that would lead to exactly \eref{TRIPLE_hnonanal} in the low-energy effective theory.
However, there do exist familiar examples where integrating out heavy degrees of freedoms yields non-analytic effective interactions.  
One is the SM plus a chiral 4th generation which, when integrated out at one loop, generates $V \supset (H^\dagger H)^2\log (H^\dagger H)$. 
Another is a model with the second Higgs doublet $\Phi$ and the potential $V_{\rm UV} =  \kappa |\Phi|^4 + \mu (\Phi^\dagger H+ \hc)$, where integrating $\Phi$ at tree level yields $V \supset (H^\dagger H)^{2/3}$.  
Yet another example is the model of Ref.~\cite{Galloway:2013dma}, which in a certain parametric limits leads to an $h$ tadpole in the effective potential, 
thus $V \supset (H^\dagger H)^{1/2}$. 
It will be clear from the following discussion that the precise form of \eref{TRIPLE_hnonanal} is not important for our argument, as long as the potential is described by a non-analytic function of $H^\dagger H$. 

For this discussion it is more convenient to work with the linear parametrization of the Higgs doublet: $H = {1 \over \sqrt 2} \bvec   i  G_1+G_2 \\  v + h + i G_3  \evec$.\footnote{%
This is because we assumed no modifications to other Higgs couplings. 
Then, in the linear parametrization, the Goldstone bosons do not have derivative couplings, which simplifies the analysis. }
Then, outside the unitary gauge, the lagrangian in \eref{TRIPLE_hnonanal} contains interactions between the Higgs and the Goldstones:  
\beq
\label{eq:TRIPLE_hnonanal2}
V \supset 
\Delta_3  {m_h^2 \over 2 v} \left (\sqrt{ (h+v)^2 + G^2} - v \right )^3 , 
\eeq 
where $G^2 \equiv  G_iG_i$,  and we do not display the Goldstone-Higgs interactions originating from the analytic SM part of the potential. 
By the equivalence theorem~\cite{Cornwall:1974km}, these correspond to interactions of longitudinal components of the $W$ and $Z$ bosons at high energies. 
This way, the non-analytic terms effectively introduce hard contact interactions between $W_L/Z_L$ and an arbitrary number of Higgs bosons. 
In particular, expanding \eref{TRIPLE_hnonanal} in $G^2$, the terms with two Goldstone boson fields are 
\bea
\label{eq:lg2}
 V & \supset &  \Delta_3 {3 m_h^2  \over 4 v} {G^2 h^2 \over h+ v} 
= \Delta_3 {3 m_h^2 \over 4} G^2  \sum_{n=2}^\infty \left (-h \over v \right )^n. 
\eea 
We can see that, for {\em any} non-zero $\Delta_3$,  \eref{lg2} contains higher-order interactions of the Higgs and Goldstone boson suppressed only by the EW scale $v$.    
It is thus clear that an EFT with the scalar potential in \eref{TRIPLE_hself} must have a low cut-off scale, $m_* \lesssim 4 \pi v$. 

The need for a UV-completion below a certain scale  manifests itself as a breakdown of perturbation theory
 around that scale. 
 This always happens because of the presence of interaction terms of dimension $>4$ in the lagrangian, carrying coefficients with negative mass dimension.
 The critical operator dimension $4$ can be overcome  by either powers of derivatives or powers of fields. In the more familiar case, like for instance 2-to-2 scattering of longitudinal vectors  in the Higgsless SM~\cite{Lee:1977eg}, the loss of perturbativity is driven by derivative interactions which make amplitudes grow with energy. In the case at hand,
like  for massive fermions in the Higgsless SM ~\cite{Maltoni:2001dc,Dicus:2004rg}, it is instead the presence of operators with an arbitrarily large number of legs that causes the breakdown of perturbation theory. 
Indeed, from \eref{lg2}, the tree-level 2-two-2 amplitude $G G \to h h$ is perfectly well-behaved and  perturbative as long as $|\Delta_3| \lesssim \cO(10)$.
In order to quantify the validity regime of an EFT with the interactions in \eref{lg2}, we have to investigate $2 \to n$ amplitudes with $n \geq 3$.

Before proceeding we would like to briefly review the logic of the standard estimates of the validity of the EFT. These are normally done by invoking the notion of breakdown of {\it perturbative unitarity}. This is conceptually fine as long as one does not interpret the issue of unitarity  too strictly. Of course there is never an issue with unitarity, as the adjective {\it perturbative} implies. The point is simply and purely the breakdown of perturbation theory associated to the onset of a strong coupling regime.
Focussing on the $S$-matrix, we know of course that unitarity is guaranteed, that is $S=e^{i\Delta}$ with $\Delta$ a Hermitian operator. The only issue concerns the ability to compute $\Delta$ in perturbation theory. $\Delta$ is a scattering phase operator, whose  eigenvalues are defined modulo $2\pi$: the  scattering phase shift is  maximized when 
an eigenvalue equals $\pi$. The regime of weak coupling can thus be defined by the request $\delta_i \lsim \pi$ for the eigenvalues of $\Delta$. Now, the computation of the $S$ matrix in perturbation theory can be phrased as a computation of $\Delta$.
In so doing unitarity is manifestly satisfied order by order in perturbation theory. Writing $S=1+iT$ we have $\Delta =-i\ln (1+iT)= T-iT^2+\dots$,
so that in the Born approximation $\Delta$ and $T$ coincide: $\Delta_0=T_0$. A rough but reasonable way to require perturbativity is thus to ask for
\beq
\langle \Psi | T_0^\dagger T_0|\Psi\rangle\lsim \pi^2
\label{NDA}
\eeq
for any incoming state $|\Psi\rangle$. Considering elastic 2-to-2 scattering one can easily check that this prescription produces  the usual NDA bounds on couplings ~\cite{Manohar:1983md,Cohen:1997rt,Luty:1997fk}. In what follows we shall simply apply this to the processed $GG\to h^n$.

Consider a family of scattering amplitudes of the isospin-0 two-Goldstone state  $| [G G]_{I=0} \rangle  \equiv  {\sum_i  |G_i G_i\rangle  \over \sqrt 3}$. 
From \eref{lg2}, the leading high-energy contribution to the inelastic amplitude for scattering this state into $n$ Higgs bosons is given by 
\beq
\label{eq:mGGtonh}
\cM([G G]_{I=0} \to \underbrace{h \dots h }_n )  \approx  (-1)^{n+1} \Delta_3 {3 \sqrt 3 n! m_h^2  \over  2 v^{n}}, 
\eeq 
and the corresponding s-wave amplitude is $\cM([G G]_{I=0}^{l=0} \to h^n)= {1 \over 4 \sqrt{\pi}} \cM([G G]_{I=0} \to h^n)$. 
A $2 \to n$ amplitude with $n>2$ that is not suppressed at large energies leads to onset of strong coupling  at some finite value  $\Lambda_*$ of the center-of-mass energy $\sqrt{s}$. 
Indeed taking $|\Psi\rangle$ to coincide with the $s$-wave $GG$ state $| [G G]_{I=0}^{\ell=0}\rangle$ the bound in eq.~(\ref{NDA})  is easily seen to read
\beq 
\label{eq:unitaritybound}
\left . \sum_{n=2}^\infty {1 \over n!}  \int  d\Pi_{n}  |\cM([G G]_{I=0}^{l=0} \to h^n)|^2 \right |_{\sqrt{s} = \Lambda_* } 
=  \sum_{n=2}^\infty  {1 \over n!} V_n(\Lambda_*)  |\cM([G G]_{I=0}^{l=0} \to h^n)|^2 
\sim \pi^2, 
\eeq 
where  $V_n(x) =   \int  d\Pi_{n}  = {x^{2n-4} \over 2 (n-1)! (n-2)! (4 \pi)^{2n-3}}$ is the volume \cite{Kleiss:1985gy} of the n-body phase space in the limit the $m_h \to 0$.\footnote{%
This approximation clearly breaks down for large enough $n$. 
However, one can show that the unitarity bounds are dominated by $n_{\rm max} \sim {\Lambda_*^2 \over (4 \pi v)^2 } \sim 4 \log \left (4 \pi v \over m_h  |\Delta_3|^{1/2} \right )$. 
For $|\Delta_3| \gtrsim 1$  we have $n_{\rm max} \sim $ few, in which case the effect of the Higgs mass on the phase space integral at high energies can be safely neglected.   
}
Inserting the explicit form of the amplitude and performing the sum over $n$, the above condition reduces to 
\beq 
   \Delta_3^2 {27 m_h^4 \over 512 \pi^2 v^4}  \left (2 + {\Lambda_*^2 \over (4 \pi v)^2 } \right )
  \exp \left (  {\Lambda_*^2 \over (4 \pi v)^2 } \right ) 
  \sim \pi^2,  
\eeq 
By definition $m_* \leq \Lambda_*$, from which we obtain the unitarity bound on the BSM scale: 
\beq
\label{eq:TRIPLE_lstar}
{m_* \over 4 \pi v } \lesssim  2 \log^{1/2} \left (4 \pi v \over m_h  |\Delta_3|^{1/2} \right ) \sim \cO(1). 
\eeq
For $|\Delta_3| \sim 1$ the maximum scale of the UV completion is parametrically of order $4 \pi v  \sim 3$~TeV, as expected.\footnote{%
We stress that the effect we discuss is unrelated to the one in \cite{Khoze:2014kka}, 
which claims the onset of strong coupling within the SM in multi-Higgs amplitudes near the production threshold. 
Our effect arises from a still perturbative contact interaction way above  threshold and is free from the subtleties existing in \cite{Khoze:2014kka} and  arising from the interplay between (large) non-perturbative amplitude and (small) phase space.
 } 
In fact, that scale is only logarithmically sensitive to the magnitude of $|\Delta_3|$, and thus remains of order  $4 \pi v$ even for $|\Delta_3| \ll 1$. 
In this bound we have only considered the $h^n$  final states. 
In reality, final states involving any number of $GG$ pairs are equally important.
Our computation thus represents a lower bound of $\langle \Psi | T_0^\dagger T_0|\Psi\rangle$, while the true upper bound on the cut-off is lower.
Further optimization of the bound  is possible by exploiting $n \to n$ scattering of special  multi-particle Higgs and Goldstone states~\cite{Chang:2019vez}.  
These improvements do not change the parametric dependence of the limit in \eref{TRIPLE_lstar}, and are not essential for our argument. 

It is clear from our argument that the bound on $m_*$ will depend little on the precise form of \eref{TRIPLE_hself}. 
A similar bound can be derived whenever the potential (or any other part of the lagrangian) contains terms non-analytic in $H^\dagger H$ that cannot be removed by field redefinitions or equations of motion. 
In such a case, higher-dimensional interaction terms between Higgs  and Goldstone bosons 
are suppressed only by powers of the EW scale $v$, leading to an onset of strong coupling in $2 \to n$ amplitudes at the scale of order $4 \pi v$. 
Such a set-up is equivalent to the SMEFT with the expansion parameter $m_* \lesssim 4 \pi v$, where gauge invariant operators with large canonical dimensions may dominate contributions to scattering amplitudes.
Only when the EFT lagrangian is analytic in $H^\dagger H$, and its terms organized as an expansion in $1/m_*$ with $m_* \gg m_h$, can the validity regime of the EFT be parametrically extended  above the EW scale.    
Such an EFT is a low-energy approximation of  BSM models with the scale separation $m_* \gg m_h$,  which were discussed in \sref{SMEFT}.

\section{Conclusions}

In this paper we derived bounds on the Higgs boson self-interactions, which are valid when the mass scale $m_*$ of BSM particles is  hierarchically larger than the EW scale. 
Under this assumption the low-energy EFT describing Higgs interactions at the EW scale is the SMEFT, organized as an expansion in $1/m_*$. 
Corrections to the cubic couplings arise at $\cO(m_*^{-2})$, that is from dimension-6 operators in the SMEFT lagrangian. 
Power counting suggests that the relative correction $\Delta_3$ to the cubic Higgs coupling can be enhanced when the BSM theory at $m_*$ is strongly coupled, such that  $\Delta_3 \gtrsim \cO(10)$ even if corrections to other Higgs couplings are $\cO(10 \%)$. 
However, this simple estimate ignores the issue of vacuum stability. 
Taking that carefully into account, we found the allowed and excluded parameter regions displayed in \fref{analytic}, which is the central result of this paper.  

Enhancement of the cubic in the range $0 \leq \Delta_3 \leq 2$ is possible under very broad assumptions. 
In particular, corrections in this range are robustly compatible with $\xi\sim 0.1$ and with $m_*$ ranging from $\sim 1$ TeV
for weak coupling 
to $\sim 5$ TeV for strong coupling. A significant portion of this region is therefore outside the present reach of LHC data.
On the other hand, outside the range $0 \leq \Delta_3 \leq 2$, vacuum stability depends on the pattern of SMEFT operators with dimensions higher than six, which in turn depends on the details of the BSM theory at the scale $m_*$.   
In view of that, it is impossible to derive sharp bounds, however, given the present experimental constraints on $\xi$, 
values $|\Delta_3|\gsim 4$ appear rather implausible, even allowing for a maximally strongly coupled BSM theory.  
Stronger limits on $\Delta_3$ hold for moderate $g_*$ or for smaller $\xi$, as visible in \fref{analytic}.   
The bottom  line is that, in the case  $|\Delta_3| \gg 1$ is measured by experiment,  we immediately learn important facts about the microscopic theory underlying the SM. 
First of all, it has to be  rather strongly coupled.
Furthermore, the parameter $\xi$ should be at least a few percent, which implies that BSM deviations in single Higgs boson couplings  may also be within the LHC reach. 
The flip side of that last statement is that improved limits on the single Higgs couplings will translate into a stronger bound on $\Delta_3$. 

It is important to stress that the upper values of $\Delta_3$, indeed $\Delta_3=\cO(1)$, can never be obtained in the more natural models of EW symmetry breaking, like  composite Higgs or  supersymmetric models. In those models, even when the Higgs is strongly coupled, there is a symmetry controlling the size of all terms in the Higgs potential. Indeed in the case of generic composite Higgs models one has $\Delta_3 \sim \xi$ like for all other Higgs couplings. 
Our scenario for maximizing $\Delta_3$  while keeping $m_*$ above the weak scale crucially relies on $m_h$ and $v$ being suppressed with respect to their natural values, $m_*$ and $f$. 
That is completely consistent, but necessarily accidental or fine tuned.  

Obviously, our bounds are not set in stone. 
There is always the possibility of a theory with either $m_*\lsim 1$ TeV or $\xi 
\sim 1$ escaping, via multiple tunings, all phenomenological constraints from Higgs and EW precision measurements and from direct searches.
 Still, given the outcome of direct BSM searches at the LHC, as well as a wide range of precision measurements that returned results consistent with the SM predictions, we believe that to be a less likely option to enhance $\Delta_3$  than our accidentally light Higgs.  
 For this reason we believe that the bounds presented in this paper are robust. 

We also investigated a more general  EFT where the Higgs potential at the EW scale cannot be written as a power series in $H^\dagger H$.
We studied corrections to the cubic Higgs self-coupling that, in a $SU(2)_W \times U(1)_Y$ invariant language,  are described by a non-analytic function of $H^\dagger H$. 
At first sight, this scenario may offer more freedom to arrange for a large $\Delta_3$ without violating stability or experimental constraints. 
We have shown however that in such a setting there is an obstruction to decoupling $m_*$ from the EW scale, leading to $\xi \sim 1$.  
Namely, $2 \to n$ amplitudes for scattering of  longitudinally polarized $W$ and $Z$ bosons into $n \geq 3$ Higgs bosons become strong and violate perturbative unitarity around the scale $4 \pi v \approx 3$~TeV.    
Therefore, in this scenario it is impossible to have a sizable $\Delta_3$ while robustly satisfying all the constraints from single Higgs processes, EW precision measurements and direct searches.
Again,  it is not completely excluded that multiple tunings and/or clever model building \cite{Galloway:2013dma,Chang:2014ida} may allow one to circumvent these phenomenological constraints.

Our analysis exemplifies the physical difference between Higgs EFTs with analytic and non-analytic potential. 
In the standard nomenclature, these EFTs go under the names of the SMEFT and the HEFT, respectively.
Previously, the distinction between the two theories was described in a less intuitive language of linearly or non-linearly realized symmetries.
Both of these EFTs have the same particle spectrum (that of the SM), however the HEFT is usually introduced as a more general theory where the $SU(2)_W \times U(1)_Y$ symmetry acts in a non-linear way on the Goldstone bosons, while the Higgs boson $h$ is an EW singlet. 
This results in more freedom in writing the Higgs potential at the leading order in the EFT expansion.    
In this paper we provided a clear and intuitive dynamical distinction between the SMEFT and the HEFT.
We argued that the HEFT can be equivalently formulated with a linearly realized  $SU(2)_W \times U(1)_Y$ symmetry, provided one allows in the lagrangian  terms that are non-analytic in $H$ around $H=0$. In our classification analyticity versus non analyticity in $H$, modulo field redefinitions, is what distinguishes SMEFT from HEFT. 
In this paper we discussed only the Higgs potential, but the same classification can be used to distinguish SMEFT vs HEFT at the level of Higgs interactions with other fields.
Our  classification is not just a matter of aesthetics and directly concerns the dynamics. Indeed the non-analyticity in $H$  makes manifest, via the equivalence theorem,  the existence of  the strong $2 \to n$ amplitudes mentioned in the previous paragraph, which prohibit extending the validity of that HEFT above the scale $4 \pi v$ \footnote{ Ref.~\cite{Alonso:2016oah} proposed another criterion to distinguish  SMEFT and HEFT. That criterion states that SMEFT corresponds to the special subclass of HEFT for which there exists a point in field space where electroweak symmetry is restored, or, equivalently, where $m_{W,Z}=0$. The resulting SMEFT class contains the SMEFT class defined by our criterion. However it seems to us it is strictly larger,  as it also includes effective lagrangians that are non-analytic at $H=0$ and thus unavoidably associated to a low cut-off scale. For instance, it seems to us that e.g. $ \cL=[1+\epsilon (H^\dagger H)^{3/2}] |D H|^2$ would be classified as SMEFT according to the criterion of Ref.~\cite{Alonso:2016oah} and as HEFT according to ours. Indeed the $n$-point amplitudes in this model, similarly to the case studied in this paper, imply a low cut-off scale, which makes our criterion appear more physical.}. Therefore, the HEFT is an appropriate low-energy description for non-decoupling BSM models with the mass scale close to a TeV.  
Conversely, BSM models with the mass scale $m_*$ parametrically larger than the EW scale are described at low energies by the SMEFT.

\section*{Acknowledgments}

We are grateful to Spencer Chang and Markus Luty  for sharing their closely related work with us, 
  and we also thank Nima Arkani-Hamed, Fabio Maltoni, Sasha Monin and Luca Vecchi for insightful discussions. 
A.F. is partially supported by the European Union's Horizon 2020 research and innovation programme under the Marie Sk\l{}odowska-Curie grant agreements No 690575 and No 674896. R.R. is partially supported by the Swiss National Science Foundation under contract 200020-169696 and through the National Center of Competence in Research SwissMAP

\bibliographystyle{JHEP} 
\bibliography{which}

\providecommand{\href}[2]{#2}\begingroup\raggedright\begin{thebibliography}{10}

\bibitem{Khachatryan:2016vau}
{\scshape ATLAS, CMS} collaboration, G.~Aad et~al., \emph{{Measurements of the
  Higgs boson production and decay rates and constraints on its couplings from
  a combined ATLAS and CMS analysis of the LHC pp collision data at $
  \sqrt{s}=7 $ and 8 TeV}},
  \href{http://dx.doi.org/10.1007/JHEP08(2016)045}{\emph{JHEP} {\bf 08} (2016)
  045}, [\href{http://arxiv.org/abs/1606.02266}{{\tt 1606.02266}}].

\bibitem{Aad:2019uzh}
{\scshape ATLAS} collaboration, G.~Aad et~al., \emph{{Combination of searches
  for Higgs boson pairs in $pp$ collisions at $\sqrt{s} = $13 TeV with the
  ATLAS detector}},  \href{http://arxiv.org/abs/1906.02025}{{\tt 1906.02025}}.

\bibitem{Sirunyan:2018two}
{\scshape CMS} collaboration, A.~M. Sirunyan et~al., \emph{{Combination of
  searches for Higgs boson pair production in proton-proton collisions at
  $\sqrt{s} = $ 13 TeV}},
  \href{http://dx.doi.org/10.1103/PhysRevLett.122.121803}{\emph{Phys. Rev.
  Lett.} {\bf 122} (2019) 121803}, [\href{http://arxiv.org/abs/1811.09689}{{\tt
  1811.09689}}].

\bibitem{McCullough:2013rea}
M.~McCullough, \emph{{An Indirect Model-Dependent Probe of the Higgs
  Self-Coupling}}, \href{http://dx.doi.org/10.1103/PhysRevD.90.015001,
  10.1103/PhysRevD.92.039903}{\emph{Phys. Rev.} {\bf D90} (2014) 015001},
  [\href{http://arxiv.org/abs/1312.3322}{{\tt 1312.3322}}].

\bibitem{Gorbahn:2016uoy}
M.~Gorbahn and U.~Haisch, \emph{{Indirect probes of the trilinear Higgs
  coupling: $gg \to h$ and $h \to \gamma \gamma$}},
  \href{http://dx.doi.org/10.1007/JHEP10(2016)094}{\emph{JHEP} {\bf 10} (2016)
  094}, [\href{http://arxiv.org/abs/1607.03773}{{\tt 1607.03773}}].

\bibitem{Degrassi:2016wml}
G.~Degrassi, P.~P. Giardino, F.~Maltoni and D.~Pagani, \emph{{Probing the Higgs
  self coupling via single Higgs production at the LHC}},
  \href{http://dx.doi.org/10.1007/JHEP12(2016)080}{\emph{JHEP} {\bf 12} (2016)
  080}, [\href{http://arxiv.org/abs/1607.04251}{{\tt 1607.04251}}].

\bibitem{DiVita:2017eyz}
S.~Di~Vita, C.~Grojean, G.~Panico, M.~Riembau and T.~Vantalon, \emph{{A global
  view on the Higgs self-coupling}},
  \href{http://dx.doi.org/10.1007/JHEP09(2017)069}{\emph{JHEP} {\bf 09} (2017)
  069}, [\href{http://arxiv.org/abs/1704.01953}{{\tt 1704.01953}}].

\bibitem{Maltoni:2017ims}
F.~Maltoni, D.~Pagani, A.~Shivaji and X.~Zhao, \emph{{Trilinear Higgs coupling
  determination via single-Higgs differential measurements at the LHC}},
  \href{http://dx.doi.org/10.1140/epjc/s10052-017-5410-8}{\emph{Eur. Phys. J.}
  {\bf C77} (2017) 887}, [\href{http://arxiv.org/abs/1709.08649}{{\tt
  1709.08649}}].

\bibitem{Degrassi:2017ucl}
G.~Degrassi, M.~Fedele and P.~P. Giardino, \emph{{Constraints on the trilinear
  Higgs self coupling from precision observables}},
  \href{http://dx.doi.org/10.1007/JHEP04(2017)155}{\emph{JHEP} {\bf 04} (2017)
  155}, [\href{http://arxiv.org/abs/1702.01737}{{\tt 1702.01737}}].

\bibitem{Kribs:2017znd}
G.~D. Kribs, A.~Maier, H.~Rzehak, M.~Spannowsky and P.~Waite,
  \emph{{Electroweak oblique parameters as a probe of the trilinear Higgs boson
  self-interaction}},
  \href{http://dx.doi.org/10.1103/PhysRevD.95.093004}{\emph{Phys. Rev.} {\bf
  D95} (2017) 093004}, [\href{http://arxiv.org/abs/1702.07678}{{\tt
  1702.07678}}].

\bibitem{Henning:2018kys}
B.~Henning, D.~Lombardo, M.~Riembau and F.~Riva, \emph{{Higgs Couplings without
  the Higgs}},
  \href{http://dx.doi.org/10.1103/PhysRevLett.123.181801}{\emph{Phys. Rev.
  Lett.} {\bf 123} (2019) 181801}, [\href{http://arxiv.org/abs/1812.09299}{{\tt
  1812.09299}}].

\bibitem{Buchmuller:1985jz}
W.~Buchmuller and D.~Wyler, \emph{{Effective Lagrangian Analysis of New
  Interactions and Flavor Conservation}},
  \href{http://dx.doi.org/10.1016/0550-3213(86)90262-2}{\emph{Nucl.Phys.} {\bf
  B268} (1986) 621--653}.

\bibitem{Leung:1984ni}
C.~N. Leung, S.~T. Love and S.~Rao, \emph{{Low-Energy Manifestations of a New
  Interaction Scale: Operator Analysis}},
  \href{http://dx.doi.org/10.1007/BF01588041}{\emph{Z. Phys.} {\bf C31} (1986)
  433}.

\bibitem{deFlorian:2016spz}
{\scshape LHC Higgs Cross Section Working Group} collaboration, D.~de~Florian
  et~al., \emph{{Handbook of LHC Higgs Cross Sections: 4. Deciphering the
  Nature of the Higgs Sector}},  \href{http://arxiv.org/abs/1610.07922}{{\tt
  1610.07922}}.

\bibitem{Contino:2017moj}
R.~Contino, D.~Greco, R.~Mahbubani, R.~Rattazzi and R.~Torre, \emph{{Precision
  Tests and Fine Tuning in Twin Higgs Models}},
  \href{http://dx.doi.org/10.1103/PhysRevD.96.095036}{\emph{Phys. Rev.} {\bf
  D96} (2017) 095036}, [\href{http://arxiv.org/abs/1702.00797}{{\tt
  1702.00797}}].

\bibitem{Glioti:2018roy}
A.~Glioti, R.~Rattazzi and L.~Vecchi, \emph{{Electroweak Baryogenesis above the
  Electroweak Scale}},
  \href{http://dx.doi.org/10.1007/JHEP04(2019)027}{\emph{JHEP} {\bf 04} (2019)
  027}, [\href{http://arxiv.org/abs/1811.11740}{{\tt 1811.11740}}].

\bibitem{DiLuzio:2017tfn}
L.~Di~Luzio, R.~Grober and M.~Spannowsky, \emph{{Maxi-sizing the trilinear
  Higgs self-coupling: how large could it be?}},
  \href{http://dx.doi.org/10.1140/epjc/s10052-017-5361-0}{\emph{Eur. Phys. J.}
  {\bf C77} (2017) 788}, [\href{http://arxiv.org/abs/1704.02311}{{\tt
  1704.02311}}].

\bibitem{Giudice:2007fh}
G.~Giudice, C.~Grojean, A.~Pomarol and R.~Rattazzi, \emph{{The
  Strongly-Interacting Light Higgs}},
  \href{http://dx.doi.org/10.1088/1126-6708/2007/06/045}{\emph{JHEP} {\bf 0706}
  (2007) 045}, [\href{http://arxiv.org/abs/hep-ph/0703164}{{\tt
  hep-ph/0703164}}].

\bibitem{Liu:2016idz}
D.~Liu, A.~Pomarol, R.~Rattazzi and F.~Riva, \emph{{Patterns of Strong Coupling
  for LHC Searches}},
  \href{http://dx.doi.org/10.1007/JHEP11(2016)141}{\emph{JHEP} {\bf 11} (2016)
  141}, [\href{http://arxiv.org/abs/1603.03064}{{\tt 1603.03064}}].

\bibitem{Contino:2016jqw}
R.~Contino, A.~Falkowski, F.~Goertz, C.~Grojean and F.~Riva, \emph{{On the
  Validity of the Effective Field Theory Approach to SM Precision Tests}},
  \href{http://dx.doi.org/10.1007/JHEP07(2016)144}{\emph{JHEP} {\bf 07} (2016)
  144}, [\href{http://arxiv.org/abs/1604.06444}{{\tt 1604.06444}}].

\bibitem{Galloway:2013dma}
J.~Galloway, M.~A. Luty, Y.~Tsai and Y.~Zhao, \emph{{Induced Electroweak
  Symmetry Breaking and Supersymmetric Naturalness}},
  \href{http://dx.doi.org/10.1103/PhysRevD.89.075003}{\emph{Phys. Rev.} {\bf
  D89} (2014) 075003}, [\href{http://arxiv.org/abs/1306.6354}{{\tt
  1306.6354}}].

\bibitem{Cornwall:1974km}
J.~M. Cornwall, D.~N. Levin and G.~Tiktopoulos, \emph{{Derivation of Gauge
  Invariance from High-Energy Unitarity Bounds on the s Matrix}},
  \href{http://dx.doi.org/10.1103/PhysRevD.10.1145,
  10.1103/PhysRevD.11.972}{\emph{Phys. Rev.} {\bf D10} (1974) 1145}.

\bibitem{Lee:1977eg}
B.~W. Lee, C.~Quigg and H.~B. Thacker, \emph{{Weak Interactions at Very
  High-Energies: The Role of the Higgs Boson Mass}},
  \href{http://dx.doi.org/10.1103/PhysRevD.16.1519}{\emph{Phys. Rev.} {\bf D16}
  (1977) 1519}.

\bibitem{Maltoni:2001dc}
F.~Maltoni, J.~M. Niczyporuk and S.~Willenbrock, \emph{{The Scale of fermion
  mass generation}},
  \href{http://dx.doi.org/10.1103/PhysRevD.65.033004}{\emph{Phys. Rev.} {\bf
  D65} (2002) 033004}, [\href{http://arxiv.org/abs/hep-ph/0106281}{{\tt
  hep-ph/0106281}}].

\bibitem{Dicus:2004rg}
D.~A. Dicus and H.-J. He, \emph{{Scales of fermion mass generation and
  electroweak symmetry breaking}},
  \href{http://dx.doi.org/10.1103/PhysRevD.71.093009}{\emph{Phys. Rev.} {\bf
  D71} (2005) 093009}, [\href{http://arxiv.org/abs/hep-ph/0409131}{{\tt
  hep-ph/0409131}}].

\bibitem{Manohar:1983md}
A.~Manohar and H.~Georgi, \emph{{Chiral Quarks and the Nonrelativistic Quark
  Model}}, \href{http://dx.doi.org/10.1016/0550-3213(84)90231-1}{\emph{Nucl.
  Phys.} {\bf B234} (1984) 189--212}.

\bibitem{Cohen:1997rt}
A.~G. Cohen, D.~B. Kaplan and A.~E. Nelson, \emph{{Counting 4 pis in strongly
  coupled supersymmetry}},
  \href{http://dx.doi.org/10.1016/S0370-2693(97)00995-7}{\emph{Phys. Lett.}
  {\bf B412} (1997) 301--308}, [\href{http://arxiv.org/abs/hep-ph/9706275}{{\tt
  hep-ph/9706275}}].

\bibitem{Luty:1997fk}
M.~A. Luty, \emph{{Naive dimensional analysis and supersymmetry}},
  \href{http://dx.doi.org/10.1103/PhysRevD.57.1531}{\emph{Phys. Rev.} {\bf D57}
  (1998) 1531--1538}, [\href{http://arxiv.org/abs/hep-ph/9706235}{{\tt
  hep-ph/9706235}}].

\bibitem{Kleiss:1985gy}
R.~Kleiss, W.~J. Stirling and S.~D. Ellis, \emph{{A New Monte Carlo Treatment
  of Multiparticle Phase Space at High-energies}},
  \href{http://dx.doi.org/10.1016/0010-4655(86)90119-0}{\emph{Comput. Phys.
  Commun.} {\bf 40} (1986) 359}.

\bibitem{Khoze:2014kka}
V.~V. Khoze, \emph{{Perturbative growth of high-multiplicity W, Z and Higgs
  production processes at high energies}},
  \href{http://dx.doi.org/10.1007/JHEP03(2015)038}{\emph{JHEP} {\bf 03} (2015)
  038}, [\href{http://arxiv.org/abs/1411.2925}{{\tt 1411.2925}}].

\bibitem{Chang:2019vez}
S.~Chang and M.~A. Luty, \emph{{The Higgs Trilinear Coupling and the Scale of
  New Physics}},  \href{http://arxiv.org/abs/1902.05556}{{\tt 1902.05556}}.

\bibitem{Chang:2014ida}
S.~Chang, J.~Galloway, M.~Luty, E.~Salvioni and Y.~Tsai, \emph{{Phenomenology
  of Induced Electroweak Symmetry Breaking}},
  \href{http://dx.doi.org/10.1007/JHEP03(2015)017}{\emph{JHEP} {\bf 03} (2015)
  017}, [\href{http://arxiv.org/abs/1411.6023}{{\tt 1411.6023}}].

\bibitem{Alonso:2016oah}
R.~Alonso, E.~E. Jenkins and A.~V. Manohar, \emph{{Geometry of the Scalar
  Sector}}, \href{http://dx.doi.org/10.1007/JHEP08(2016)101}{\emph{JHEP} {\bf
  08} (2016) 101}, [\href{http://arxiv.org/abs/1605.03602}{{\tt 1605.03602}}].

\end{thebibliography}\endgroup

\end{document}